# A Two-Level Thermal Cycling-aware Task Mapping Technique for Reliability Management in Manycore Systems

**Fatemeh Hossein-Khani[1], Omid Akbari[1], and Muhammad Shafique[2], Senior Member, IEEE**

[1]Department of Electrical and Computer Engineering, Tarbiat Modares University, Tehran 14115-111, Iran
[2]Division of Engineering, New York University Abu Dhabi (NYU AD), Abu Dhabi 129188, United Arab Emirates

Corresponding author: Omid Akbari (e-mail: o.akbari@modares.ac.ir).

This work has been supported in parts by the NYUAD Center for Cyber Security (CCS), funded by Tamkeen under the NYUAD Research Institute Award G1104.

**ABSTRACT** Reliability management is one of the primary concerns in manycore systems design. Different aging mechanisms such as Negative-Bias Temperature Instability (NBTI), Electromigration (EM), and thermal cycling (TC) can reduce the reliability of these systems. However, state-of-the-art works mainly focused on NBTI and EM, whereas a few works have considered the thermal cycling effect. The thermal cycling effect can significantly aggravate the system's lifetime. Moreover, the thermal effects of cores on each other due to their adjacency may also influence the system's Mean Time to Failure (MTTF). This paper introduces a new technique to manage the reliability of manycore systems. The technique considers thermal cycling, adjacency of cores, and process variation-induced diversity of operating frequencies. It uses two levels of task mapping to improve system lifetime. At the first level, cores with close temperatures are packed into the same bin, and then, an arrived task is assigned to a bin with a similar temperature. Afterward in the second level, the task is assigned to a core inside the selected bin in the first level, based on performance requirements and the core frequency. Compared to the conventional Thermal cycling aware techniques, the proposed method is performed at a higher level (bins level) to reduce the thermal variations of cores inside a bin, and improves the system MTTF$_{TC}$, making it a promising solution for manycore systems. The efficacy of our proposed technique is evaluated on 16, 32, 64, and 256 core systems using SPLASH2 and PARSEC benchmark suite applications. The results show up to 20% MTTF$_{TC}$ increment compared to the conventional thermal cycling-aware task mapping techniques.

## I. INTRODUCTION

Technology advancements in the nanoscale era allow the integration of a higher number of cores on a chip. However, reducing the operating voltage, as well as decreasing the capacitance of the nodes, make the integrated circuits (ICs) more vulnerable to the different aging mechanisms such as negative bias temperature instability (NBTI), electromigration (EM), and thermal cycling (TC). These factors may lead to timing errors and, in some cases, system failure [1]. In the following subsections, we will delve deeper into this challenge by first identifying the specific aging mechanisms that are most relevant to our target problem. We will then review the limitations of existing state-of-the-art techniques. Finally, we will present our novel approach.

### A. TARGET RESEARCH PROBLEM

Thermal cycling (TC) is a phenomenon that occurs when an electronic device or a part of it oscillates between temperatures over its lifetime [1]. As an example, TC can cause an integrated circuit (IC) to expand and contract, which can lead to mechanical stress and ultimately lead to failure. Mitigating the impacts of TC in digital circuits, as well as the manycore systems is a research venue to improve their lifetime reliability [1][2]. Previous research has shown that thermal cycling can cause mechanical failure in the systems [3]. Additionally, it can impact other aging mechanisms such as negative bias temperature instability (NBTI) and electromigration (EM). However, thermal cycling is the only aging mechanism that explicitly considers both temperature variations and temperature level [4]. Due to these characteristics of the thermal cycling mechanism, we focused on its impact on the Mean Time To Failure (MTTF) of systems in our paper, as other previous works [5][6]. Moreover, increasing computational requirements in different application domains [7], higher circuit temperature, and thus, stringent thermal design power (TDP) constraints are the major driving forces to seek more efficient power and reliability management methods in manycore systems [8].

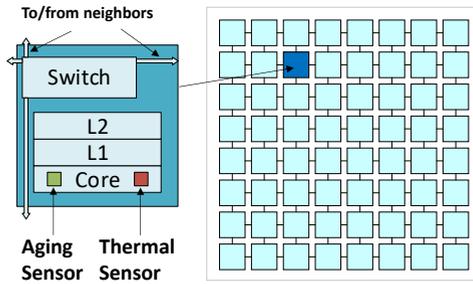

**FIGURE 1.** The architectural view of a manycore system.

The architectural view of a manycore system is shown in FIGURE 1. It is composed of a mesh of tiles, in which each tile includes a core, two levels of private L1 and L2 caches, and a switch to communicate with the neighbor cores. Additionally, each tile includes a thermal and an aging sensor to monitor the state of the core. In such a manycore system, a global controller manages the operation of cores [9]. Note that the aging sensor of each core shown in FIGURE 1 continuously monitors the performance and reliability of the cores, and provides feedback on the aging status of each core that can be used to detect and predict potential failures or performance degradation in advance. However, in our work, we need to predict and simulate the long-term effects of aging on manycore systems, which is possible by employing aging estimation models (see equations (5) and (6) to calculate the MTTF$_{TC}$ of the cores).

### B. RELEVANT STATE-OF-THE-ART AND THEIR KEY LIMITATIONS

Generally, addressing the aging and reliability challenges in a manycore system while meeting power and performance requirements is considered as a complex problem [9]. Several approaches have been proposed to increase the reliability of manycore systems, such as Dynamic Voltage and Frequency Scaling (DVFS) [10][11], and Dynamic Thermal Management (DTM) [12][13]. However, most of these works have focused on the NBTI and EM mechanisms of the aging phenomena, while only a few works have also considered the thermal cycling mechanism in their reliability management method. Nevertheless, the heat exchange between a core and its neighbors can also intensify their aging, which may considerably decrease the reliability of the system. Thus, the cores adjacency effect on the aging may also be considered during the thermal cycling management process of a manycore system.

### C. OUR NOVEL CONTRIBUTIONS AND CONCEPT OVERVIEW

In this paper, Firstly, we consider the effects of cores adjacency on the cores' thermal variations. In existing approaches, each core is assumed to be independent. However, in reality, adjacent cores can have a significant impact on each other's temperatures due to their close proximity and shared resources, such as power and cooling. This can result in increased thermal stress and temperature variation, which can have a significant effect on the reliability of the system. Thus, in this work, we consider the effects of cores adjacencies on their thermal variations, and their effect on the thermal cycling of cores. In our proposed method, the core idea to consider the heat exchange between a core and its neighbors is to use a temperature-aware bin-packing algorithm that packs cores into bins based on their temperature using the Density-Based Spatial Clustering of Applications with Noise (DBSCAN) clustering method. After the bin packing step, the task is assigned to a bin with the closest temperature to its final temperature, which reduces the thermal variation in that core and its neighbors. Note that assigning a task to a core of a given bin will have the minimum effect on the neighbors of that core since their temperature is close to the final temperature of that core after executing the task. Thus, besides reducing the thermal cycling effect of the core running the task, we also reduce the temperature variations in its neighbors. Accordingly, we propose a two-level reliability management technique for manycores, which enhance the system MTTF$_{TC}$ by mapping the task into the cores considering the thermal variations along with the cores adjacency effect on the thermal cycling. At the first level, according to the temperatures of the cores, the cores are packed into several bins, where each bin is a logical grouping of cores that have a close temperature to each other. For this step, we employ the DBSCAN method, in which cores with more close temperatures are packed into the same bins. Next, the task is assigned to a bin with the closest average temperature to its temperature (see FIGURE 2).

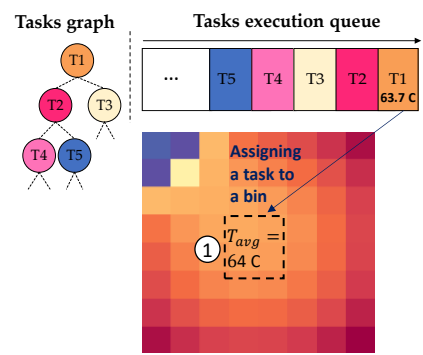

**FIGURE 2.** An example of mapping a task from the application task graph into the bins considering the thermal variations along with the core's adjacency effect on the thermal cycling. Key observations: ① A bin composed of four cores with a closest temperature to the arrived task is selected, which may result in a lower thermal variation.

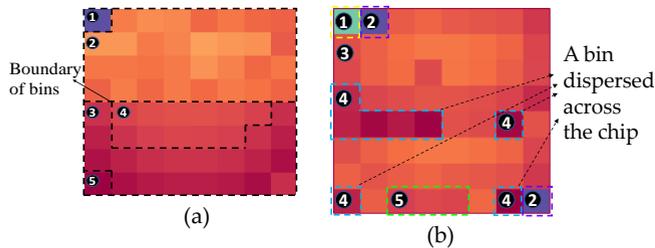

**FIGURE 3.** Two examples of five composed bins (a) when cores inside the bins are spatially contiguous, (b) when the cores of bins 2 and 4 are dispersed across the chip.

When there is more than one bin for mapping a task, we consider the distance between the cores that execute consecutive tasks, to increase the performance and lower the power consumption. Note that cores in a bin can be spatially contiguous or can be dispersed across the chip (see FIGURE 3.(a) and (b)). In the second level, the task is mapped to a core inside of the selected bin at the previous step. As mentioned before, the maximum frequency of cores may vary due to the process variation. Thus, in each bin, we assign the task with the higher performance requirements (power consumption) to the core with a higher frequency. To find the efficiency of our technique, we have provided some motivational analysis in the next subsection.

**Our novel contributions in a nutshell are:**

(1) A two-level task mapping technique to mitigate the aging process induced by the thermal cycling, as well as the process variation effects on the frequency of cores.

(2) A temperature-aware bin packing method for composing the bins based on the temperature of cores, to reduce high amplitude temperature variations and heat exchange between the cores.

(3) Considering the effects of cores adjacencies on their thermal variations, and their effect on the thermal cycling of cores. Also, packing the cores into bins using the DBSCAN method to mitigate the heat exchange of adjacent cores on each other.

(4) Studying the effects of different aging mechanisms, including thermal cycling (TC), negative bias temperature instability (NBTI), hot carrier injection (HCI), and electromigration (EM) on the MTTF of manycore systems, and how our proposed approach can mitigate the effects of those mechanism.

(5) Evaluating the proposed technique for different SPLASH2 and PARSEC benchmark suites for the various number of cores and tasks.

### D. MOTIVATIONAL CASE STUDY

As mentioned in previous subsections, the interaction and exchange of heat between a core and its neighboring cores can exacerbate their aging process. This phenomenon occurs because the temperature variations resulting from heat exchange induce higher thermal variations, ultimately diminishing the longevity of the cores. To assess the implications of heat exchange on cores reliability, we conducted a study using a 64K point Fast Fourier Transform (FFT) application from the SPLASH2 benchmark suite [14].

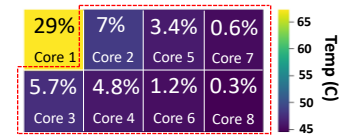

**FIGURE 4.** $MTTF_{TC}$ reduction of the free cores due to the heat exchange with the core performing FFT application.

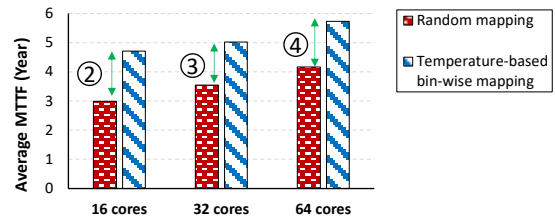

**FIGRUE 5.** $MTTF_{TC}$ of the 16, 32, and 64-core systems under the two different mapping schemes; random mapping and temperature-based bin-wise mapping techniques. Key observations: $MTTF_{TC}$ improvement in bin-wise mapping considering cores adjacency effect on each other in ② is 53%, in ③ 42%, and in ④ 45%.

This application was executed on a specific core within an eight-core system, where we monitored the resultant temperature changes in the remaining seven cores that do not perform any task. Note that the used models for this study, as well as the employed simulation tool flow will be presented in Section IV and Section V, respectively.

FIGURE 4 depicted a result of this examination, where running the 64K point FFT on core 1 is led to temperature changes of 22.3°C, 3.5°C, and 0.53°C in the core 1, core 2, and core 8, respectively. These temperature variations resulted in the TC related MTTF (i.e., $MTTF_{TC}$) reduction of the affected cores by 29%, 7%, and 0.3%, respectively. Therefore, considering the heat exchange of neighbor cores, in the task mapping step can significantly increase cores' MTTF. To mitigate the effects of these heat exchanges, in this paper, we propose a temperature-based bin packing technique, in which, at first, cores are packed into bins based on their temperature, and then, the arrived task is assigned to a bin with the closest average temperature to the final temperature of the core that will perform the task. Noe that the packing step is performed using the DPSCAN algorithm (details of this algorithm is presented in Section IV), with the goal of minimizing heat exchange between cores and improve their overall thermal stability, ultimately enhancing their reliability.

As an example of applying the aforementioned method in the task mapping step, FIGRUE 5 presents the $MTTF_{TC}$ of different systems with and without using the temperature-based bin packing. The results demonstrate that temperature-based bin packing significantly improves $MTTF_{TC}$, achieving, on average, 23%, 51%, and 40% improvement for 16, 32, and 64-core systems, respectively.

### E. MAIN RESULTS COMPARED TO STATE-OF-THE-ART

We conducted a comparative analysis of our proposed method against three other methods: the conventional TC-aware task mapping technique, the method proposed in [1], and the random task mapping method. The results demonstrated that our method achieved an up to 53% enhancement in MTTF$_{TC}$ compared to the random task mapping method.

**Paper Organization:** The rest of the paper is organized as follows: Section II presents Related works on reliability management. Section III describes the Background of our proposed technique. Our structure is presented in Section IV. Section V explains the Simulation toolchain to run our two-level task mapping technique and evaluation the technique. Finally, the conclusions of this study are discussed in Section VI.

## II. RELATED WORKS

In this section, we review prior research on manycore systems, specifically, the works that focused on increasing the lifetime of these systems, considering the thermal cycling phenomenon.

DVFS is a well-known technique that can improve the lifetime of the system since aging is a strong function of the supply voltage and temperature [10][11]. Pourmohseni et. Al., in [15] presented a task migration policy aimed at optimizing system performance, complemented by a DVFS-based approach for efficient thermal management. A multi-level thermal stress-aware power and temperature management approach was proposed in [2]. In this paper, Iranfar er. al., used the DVFS for avoiding high spatial and temporal thermal variations among the multiprocessor system-on-chip (SoC). Taking into consideration the workload-dependent nature of voltage/frequency settings for individual cores, [16] utilized DPM policies to regulate the Voltage/Frequency (VF) knobs of the cores. They further proposed a Learning-to-Search framework, leveraging these DPM policies, to achieve efficient management of power, performance, and thermal considerations. In [17], Mohammed and his colleagues introduced a dynamic thermal-aware technique called DTaPO. DTaPO aims to optimize performance on manycores by utilizing task migrations and DVFS methods as Dynamic Thermal Management (DTM) techniques. It specifically focuses on optimizing performance in dark silicon manycore systems, considering both active and dark cores. Additionally, in [18] Mohammed et. al., proposed a Dynamic Neighbor-Aware Performance Enhancement (3D-DNaPE) technique. 3D-DNaPE consists of a performance enhancement algorithm, a neighbor-aware pattern algorithm, and a DTM method that incorporates DVFS and task migration. The technique operates in two stages: first, identifying the coldest core among neighbors, and then applying DTM methods to thermally constrained 3D many-core systems.

In [1], Haghbayan et. Al., proposed a thermal-cycling-aware runtime resource management approach for manycore systems. A dynamic process variation (PV) and aging-aware mapping method was proposed in [20] by Rathore et. al., to maximize the lifetime reliability of manycore systems while meeting constraints (performance, power, and thermal). In [12], Gnad et. al., have proposed a run-time aging management method to minimize the chip aging in on-chip multi-core systems. This method considers both the Dark Silicon constraint and variation to increase the performance for a given lifetime. Task mapping in manycore systems considering performance, power consumption, and delay constraints is an NP-hard problem [13]. In [8], Rathoe et. al., have proposed a distributed resource management method to achieve maximum performance of manycores while meeting temperature constraints. Silva and other authors In [13], applied a proportional, integral, and derivative temperature management (PIDTM), using the DVFS to NoC-based systems to minimize peak temperature. Moreover, [19] Lin et. al., introduced MRDF, a cutting-edge multi-agent reinforcement learning-based approach that seamlessly integrates Dynamic Voltage Frequency Scaling (DVFS) and dynamic fan control. By effectively managing power consumption and performance, while maintaining adherence to thermal constraints, this method achieves an optimal trade-off in system operation.

In [21], an approximate computing-based method was proposed to achieve a reliable DTM, in which the accuracy of the thermal profile delivered to the DTM is considered the efficiency and reliability factor of DTM by Rahimipour and other authors. In this method, to achieve a reliable DTM, the parity method has been used for the most significant bits (MSBs) of thermal data, while there is no protection for its three least significant bits (LSBs). Das and their colleagues in [22] were focused on thermal management in multicore systems by controlling the peak and average temperatures, as well as the thermal cycling concern. To this end, the paper proposes a reinforcement learning-based approach, which learns an optimal thermal management policy that maximizes the lifetime reliability of the system while minimizing thermal stress. Specifically, the approach used a Q-learning algorithm to learn thermal management policies that optimize the temperature distribution and avoid hotspots. The algorithm also considers both inter- and intra-application thermal interactions, allowing for more effective and adaptive thermal management. One advantage of this work is its ability to adapt to changing workloads and thermal conditions, as the Q-learning algorithm can continuously learn and update the thermal management policies based on feedback from the system. However, this approach requires more computational resources and training time. Moreover, the heat exchange between cores and its related thermal variations was not considered in [22].

In [23], Latibari et. al., proposed a task scheduling and mapping method to improve the reliability of a manycore system by keeping the power of cores below the thermal safe power (TSP), where the TSP is a core-level power constraint. In this method, at first, the TSP is checked; if the task meets the TSP constraint, it will be executed. Otherwise, the task execution will shift to the next time slot.

A task mapping method was proposed in [20] by Rathore et. al., to improve the manycore system lifetime. This strategy uses application characteristics, PV, and cores aging information to determine the mapping of dark cores considering performance, power, and temperature constraints, where the dark cores are denoted to the power-gated ones due to power constraints. In [8], Reinforcement Learning (RL) was used to learn the cores aging behavior. Afterward, based on the reward and action table updated during the learning phase, tasks are mapped into the most appropriate cores that satisfy the performance requirements. A thermal stress-aware variable power and thermal management (TSA-VPTM) framework was introduced in [24] by Kamal and other authors. In this framework, the consolidation and deconsolidation-based method was used to reduce the power consumption and thermal gradients. In [25], based on theoretical results, Chantem et. al., have proposed a dynamic task assignment and scheduling method to optimize multicores' lifetime, which considers the core wearing-out state before task assignment and scheduling decisions. Also, a data distillation method has been suggested to reduce the size of thermal profiles, which allows online reliability analysis of the full system since the runtime of their reliability modeling tool depends on the size of thermal traces. As claimed in [25], wearing-out due to TDDB, EM, and stress migration (SM) primarily depends on temperature, while the TC phenomenon depends on the frequency and amplitude of thermal cycles.

In [2], a thermal stress-aware power and thermal management approach for MPSoCs was proposed. In this method, core consolidation and deconsolidation are performed, where the peak temperature and thermal stress constraints are considered to determine the optimal frequencies of cores by solving a convex optimization problem. In [1], a thermal-cycling-aware reliability management method has been proposed. This method has four different steps: a reliability analysis step that finds cores aging status, a reliability-aware mapping step with a mapping policy considering a reliability-aware affinity metric (RAF), a reliability-aware thread scheduling step, and a reliability-aware DPM, which dynamically adjust the voltage and frequency to satisfy the TDP constraint. A dynamic reliability management (DRM) method was introduced in [26] by Baldassari and other authoers. In this method, the aging status of cores are monitored, and a controller manages the resources based on the throughput and performance requirements. A dynamic reliability management (DRM) method for multiprocessors has been proposed in [27] by Srinivasan et. al., creates a tread-off between power, performance, and reliability. Instead of overly conservative approaches that design the system for the worst-case operating situations, this paper proposed a less-than-worst-case temperature qualification method to reduce the cost, without sacrificing the performance. Especially, the lifetime reliability of multiprocessors is tracked depending on the application's behavior, and then, if the reliability is retained, the application is run at higher performance, otherwise, the performance is degraded. To address the reliability concerns of soft real-time embedded systems executing on integrated CPU and GPU platforms, Zhou et. al., proposed a hybrid method in [28], which improves the soft-error reliability while retaining the lifetime reliability constraint. This method leverages various knobs, such as reducing the utilization of cores, task migration at runtime, and dynamic frequency scaling to resolve the aforementioned concerns.

There are several works that consider tasks to be independent, i.e., each task can be executed separately without relying on the completion or output of other tasks. In [29], Chakraborty et. al., proposed a method for real-time task scheduling. In this work, independent tasks are assigned to cores based on their thermal status while considering core frequencies. Similarly, [30] and [31] also consider task scheduling in homogeneous systems. In [30], Moulik and other authors introduced a temperature-aware heuristic task scheduling method which is low-overhead three-level hierarchical temperature-aware semi-partitioned proportional fair scheduler, while in [31] Sharma et. al., proposed a real-time temperature-aware task scheduling method called RESTORE that allocates tasks to cores and schedules them while meeting performance requirements using voltage/frequency scaling.

However, heterogeneous multicore systems offer different advantages such as improved performance and power efficiency compared to homogeneous systems. The utilization of diverse processor architectures, specialized accelerators, and varying computational resources in heterogeneous multicore systems can lead to enhanced performance, power efficiency, and overall system capabilities. In [32], Sharma et. al., introduced the RT-SEAT method, which is a hybrid method for real-time task scheduling while meeting performance requirements. RT-SEAT proposes a task scheduling technique that reduces power consumption of the system using VF-scaling while considering task deadlines. The term hybrid refers to the method employed in the system, where the cores are segregated into independent groups based on predefined criteria. Each task is then assigned to a specific group and is exclusively allowed to execute and migrate within the cores belonging to that particular group. In [33] Moulik et. al., proposed a task scheduling technique called CEAT, which is an energy-aware task scheduling technique for heterogeneous systems. CEAT employs DVFS to meet required performance levels and also manages the power consumption. Furthermore, in [34] Sharma and other authors proposed a fault-tolerant real-time task scheduling method named as FATS-2TC. This work emphasizes the simultaneous consideration of fault-tolerance and energy efficiency, which sets it apart from other existing works in the field. Also, Sharma et. al., proposed a real-time task scheduling method for periodic tasks involving four steps: deadline partitioning, core clustering, temperature-aware scheduling, and energy-aware scheduling in [35].

As discussed in Section I, different aging mechanisms, such as TC, NBTI, HCI, and EM may have a profound effect on the reliability of a core and its neighbors in a manycore system. In the proposed two-level mapping algorithm, to mitigate the effects of these aging mechanisms and reduce the heat exchange due to the thermal variation of a core on its neighbors, and consequently, achieve higher lifetime reliability in manycores, we considered the heat exchange of a core on its neighbors, which is a significant limitation of prior works that mitigated the thermal cycling effects.

TABLE 1
A CATEGORIZATION OF THE STUDIED STATE-OF-THE-ART
WORKS BASED ON THEIR TECHNIQUE

| Technique | Related works | Brief description |
|---|---|---|
| Voltage and frequency scaling | [10][11][16] | The voltage and frequency of cores are dynamically scaled to meet power constraints. |
| Thermal management techniques | [12]-[2][21][24] | DTM is used as a solution to avoid high spatial and temporal thermal variations. |
| Aging aware techniques | [1][8][23][26] | These works have focused on reducing the effects of one or some aging mechanisms, including NBTI, EM, and thermal cycling. |
| Process variation aware techniques | [8][20] | Process variation is considered for task mapping in manycore systems to retain the desired reliability level. |

TABLE 1 shows a categorization of studied related works based on their used techniques. As discussed, a few works have considered thermal cycling for improving system reliability. However, the effect of heat exchange due to the cores adjacencies can significantly reduce the MTTF$_{TC}$. In this work, we propose a task mapping technique to increase the system reliability by considering the thermal cycling and heat exchange due to the adjacency of cores, and the process variation effects as well.

## III. BACKGROUND
In this section, the PV and reliability models of the thermal cycling phenomenon in a manycore system are discussed.

### A. PROCESS VARIATION (PV) MODEL
In this subsection, inspired from [36], we introduce a process variation (PV) model to consider the manufacturing process variations in the simulated manycore systems. With the technology scaling in the nanoscale era, process variation can undermine the reliability of SoCs since it may influence systems physical parameters such as wire width and height, and the resistance of the power network. Especially, in manycore systems, it can change the cores characteristics and affect the safe operating frequency of cores and leakage power [8]. For modeling the PV, the chip surface is modeled as a grid with the dimension of P × Q [9].

Let $p_{(x,y)}$,($x \in [1,P]$,$y \in [1,Q]$) represent the value of the process variation information at grid cell $(x,y)$, where $(x,y)$ denotes the position of a core in the manycore system. The physical parameters of the manycore system at the grid cell $(x,y)$ are given by:

$$W_{(x,y)} = \kappa_1 p_{(x,y)} \qquad (1)$$

$$H_{(x,y)} = \kappa_2 p_{(x,y)} \qquad (2)$$

$$Res_{(x,y)} = \gamma p_{(x,y)} \qquad (3)$$

where $W_{(x,y)}$, $H_{(x,y)}$ and $Res_{(x,y)}$ are wire width, height, and power grid resistance at the cell $(x,y)$, respectively. Also, $\kappa_1$, $\kappa_2$, and $\gamma$ are technology-specific constants [9]. Finally, the maximum frequency of a core affected by the PV obtained by:

$$f_{(x,y)} = \beta \min_{s,t \in S_{(CP,x,y)}} p_{(s,t)} \qquad (4)$$

where, $S_{(CP,x,y)}$ is the set of grid cells and $\beta$ is a technology-specific constant. Also, $p_{(s,t)}$ is the process variation information at grid point $(s,t)$.

The introduced PV model estimates the core frequencies based on the statistical distribution through (1)~(3), i.e., the operating frequency of each core is estimated regarding the PV model.

### B. RELIABILITY MODEL
Different temperature-related aging mechanisms such as negative bias temperature instability (NBTI), hot carrier injection (HCI), and thermal cycling (TC) may affect the lifetime of a manycore system. In the following, these aging mechanisms and the corresponding MTTF models are studied.

*Thermal Cycling MTTF Model:*
The thermal cycling phenomenon occurs when the temperature rises or drops down and then goes back [1].

These cycles are counted by Downing's simple rain flow-counting algorithm proposed in [37] that can be used to calculate the MTTF. However, when the number of cycles and their amplitude grows, system aging is intensified and the MTTF is reduced [1]. Note that, each cycle may include several thermal variations with different amplitudes. The Coffin-Manson equation is used to calculate the number of cycles (N$_{TC}$) defined by [1]:

$$N_{TC}(i) = A_{TC}(\delta T_i - T_{th})^{-b} exp\left(\frac{E_{aTc}}{KT_{max}}\right) \qquad (5)$$

where $A_{TC}$ and b are the empirically and coffin-Monson exponent constants, respectively. $T_{th}$ is the threshold temperature where the inelastic deformation starts, and $\delta T_i$ is the maximum thermal amplitude change of $i^{th}$ thermal cycle. Also, $E_{aTc}$ and $T_{max}$ are the activation energy and the maximum temperature during the cycle, respectively. Based on (5), MTTF is calculated by [1]:

$$MTTF_{TC} = \frac{N_{TC}\sum_{i=0}^{m} t_i}{m} \qquad (6)$$

where $t_i$ and $m$ are the duration and total number of cycles, respectively.

Note that the cycles are calculated after executing an application, to obtain required parameters (e.g., N$_{TC}$ which is the number of cycles) for calculating the MTTF.

*NBTI MTTF Model:*
Negative Bias Temperature Instability (NBTI) is a phenomenon that affects the reliability of transistors in electronic devices. NBTI is caused by the gradual buildup of charge traps in the gate oxide of a transistor when it is subjected to a negative bias voltage over time.

TABLE 2
USED NOTATIONS

| Notations used in formulas | |
| --- | --- |
| $W_{x,y}$ | Wire width |
| $H_{x,y}$ | Wire Height |
| $Res_{x,y}$ | Power grid resistance at the tile (x,y) |
| $\kappa_1, \kappa_2$, and $\gamma$ | Technology-specific constants |
| $f_{x,y}$ | Maximum frequency |
| $S_{CP,x,y}$ | Set of grid points |
| $\beta$ | Technology-specific constant |
| $N_{TC}$ | Number of cycles |
| $A_{TC}$ | Empirically constant |
| $b$ | Coffin-Monson exponent constant |
| $T_{th}$ | Threshold Temperature |
| $\delta T_i$ | Maximum thermal amplitude change in the $i^{th}$ thermal cycle. |
| $E_{aTc}$ | Activation energy |
| $T_{max}$ | Maximum temperature during the cycle |
| $t_i$ | Cycle duration |
| $m$ | Total number of cycles |
| $M_{dist}$ | Manhattan distance between two cores |
| A, B, C, D, and $\beta$ | Fitting Parameters |
| $I_{sub}$ | Peak substrate current |
| w | Width of transistor |
| Q | Activation energy |
| I | Current flowing into or out of the contact window. |

| Notations of algorithms | |
| --- | --- |
| Epsilon | Maximum difference between the temperature of cores |
| M | Minimum number of cores required to form a bin |
| $\{T_1, .., T_n\}$ | Set of cores temperature |
| $\{C_1, ..., C_n\}$ | Set of cores |
| $\{C_{f1}, ..., C_{fn}\}$ | Set of free cores |
| $\{F_1, ..., F_n\}$ | Set of cores frequencies |
| $\{F_{t1}, ..., F_{tk}\}$ | Set of tasks frequency requirements |
| $\{B_b, b \in \{1, ..., n_{Bin}\}\}$ | Formed Bins |
| $\{b_1, ..., b_{Bin\ number}\}$ | Set of free bins |

As the temperature increases, the rate of charge trapping and de-trapping in the gate oxide also increases, which can the NBTI degradation process, i.e., the MTTF affected by NBTI will decrease as the temperature increases. The MTTF change due to the NBTI phenomenon is obtained by [38]:

$$MTTF_{NBTI} \propto ([(ln\left(\frac{A}{(1+2e^{\frac{B}{KT}})}\right) - ln\left(\frac{A}{(1+2e^{\frac{B}{KT}})} - c\right)] \times \frac{T}{e^{\frac{-D}{KT}}})^{\frac{1}{\beta}} \quad (7)$$

where *A, B, C, D*, and $\beta$ are fitting parameters.

*HCI MTTF model:*
Hot Carrier Injection (HCI) is an aging mechanism that can impact the MTTF of a system, which occurs when high-energy electrons (hot carriers) are generated in the channel region of MOSFETs due to the high electric field. These hot carriers can gain enough energy to penetrate the gate oxide and become trapped there, leading to a modification of the oxide charge. The MTTF change due to the HCI mechanism is defined by [39]:

$$MTTF_{HCI} \propto (\frac{I_{sub}}{W})^{-n} \times e^{\frac{Q}{KT}} \quad (8)$$

where $I_{sub}$ is the peak substrate current, *W* is the width of transistor, and *Q* is the activation energy.

*EM MTTF model:*
Electromigration (EM) is caused by the migration of metal atoms due to momentum transfer between the electrons and metal ions in the interconnects. This migration can lead to the formation of voids or hillocks in the interconnects, which can degrade the performance and reliability of the electronic system. The MTTF model of the HCI mechanism is expressed by [39] as follow:

$$MTTF_{EM} \propto I^{-n} \times e^{\frac{Q}{KT}} \quad (9)$$

where, *I* is the current flowing into or out of the contact window, n is an empirical parameter with a value between 1 and 2, and *Q* is the electro-migration activation energy.

## IV. PROPOSED TWO-LEVEL MAPPING TECHNIQUE
In this section, first, the used manycore system and application models are described. Afterward, the proposed two-level thermal-cycling-aware task mapping technique for manycore systems is introduced. The system overview diagram of the proposed mapping technique is shown in FIGURE 6, and details of different key components of the technique are discussed in the subsequent sections.

### A. MANYCORE SYSTEM MODEL
In this work, a mesh-based manycore architecture integrating homogeneous cores connected through a Network-on-Chip (NoC) is employed (see FIGURE 1), in which per-core thermal and aging sensors are included. We demonstrate cores set by $\{C_1, C_2, ..., C_n\}$, where each core has its own maximum operating frequency given by $\{f_1, f_2, ..., f_n\}$. Note that, the differences between the cores maximum frequency are due to the process variation phenomenon discussed in Section III.

### B. APPLICATION MODEL
For the workload model, we considered multi-thread applications $A = \{A_1, A_2, ..., A_n\}$, where each application Notations used in formulas $A_p$ has multiple threads $\{t_{p,1}, t_{p,2}, ..., t_{p,n}\}$. These threads are the tasks that should be executed on the cores.

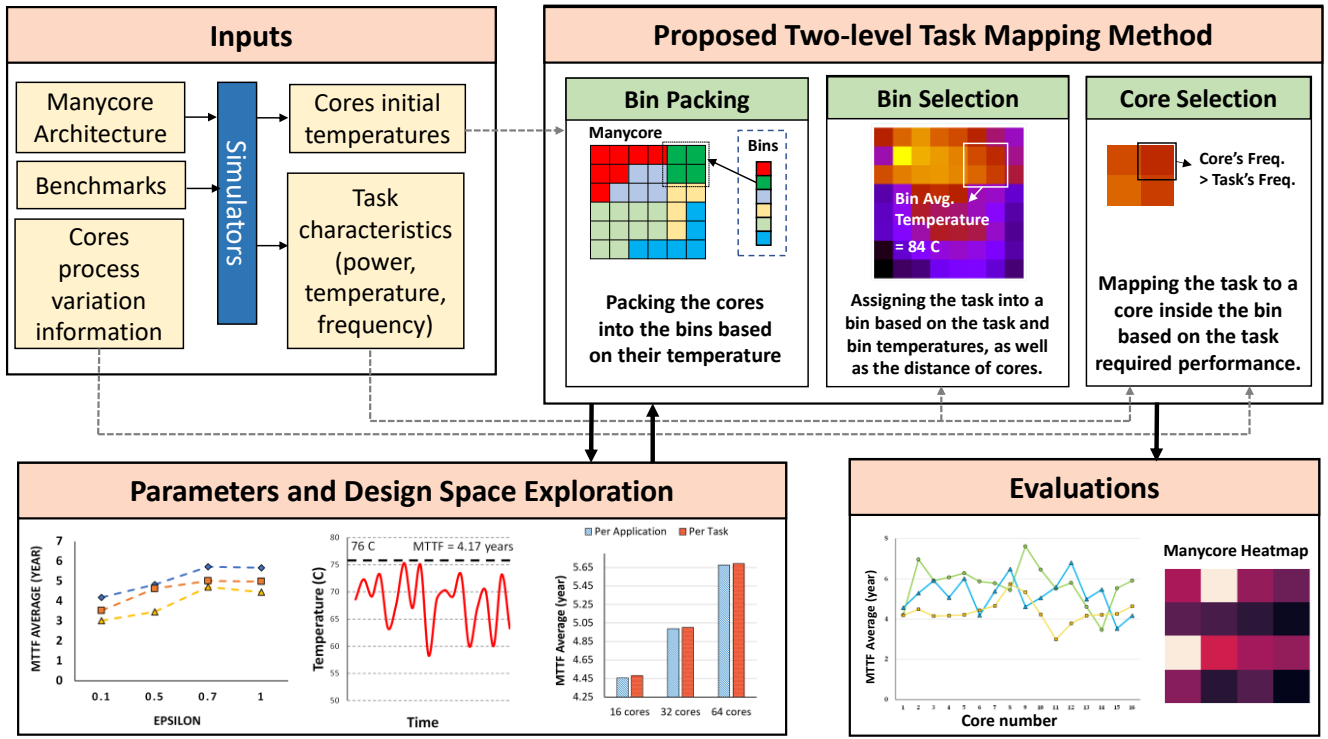

**FIGURE 6.** System overview diagram of the proposed mapping technique.

---

**Algorithm 1: Bin Packing**

**Input**: Set of cores $C = \{C_1, \ldots, C_n\}$, Set of cores temperature $T = \{T_1, \ldots, T_n\}$
**Output**: Formed bins, i.e., $\{B_b, b \in \{1, \ldots, n_{Bin}\}\}$
**Parameters**: *Epsilon, M*

1    DBSCAN($T$, *Epsilon, M*) {
2        **while**(there is a core without label)
3          choose a random core $C_r$
4          **for** c in $C$
5            **if** $|T_c - T_r| < Epsilon$
6              add outlier label to c
7            **else**
8              add noise label to c
9          **if** outliers $\geq M$
10            pack outliers and $C_r$ to a bin
11            add the bin to Formed bins
12        **return** Formed bins

---

### C. PROPOSED TWO-LEVEL TASK MAPPING

FIGURE 6 shows the overview of our proposed reliability-aware task mapping technique composed of three steps: bin packing, task-to-bin assignment, and task-to-core mapping. As shown in this figure, application threads, manycore system architecture, the temperature of cores, and process variation information is fed as the inputs to our mapping algorithm Similar to some state-of-the-art works that mitigate the thermal cycling effects (e.g., see [2][10]), we perform applications offline to extract the required parameters for the proposed mapping approach. Specifically, we perform each application once to extract temperatures of the tasks, required for the bin packing step. Then, during the runtime when applications are executing, our proposed approach utilizes offline-calculated Task temperatures, as well as online-measured cores temperatures and bin formations, to improve the reliability and lifetime of the system. Similar to [8] we considered tasks temperatures as steady-state temperature of the core that task execution will cause. Steady-state temperature refers to the maximum temperature reached by a core over time, considering the cumulative impact of all instances running on that core. Our proposed mapping steps are discussed in the following.

**Bin packing:**
To consider the thermal effects of adjacent cores, first, we pack the cores into different bins based on their temperature. Algorithm 1 explains our proposed temperature-aware bin-packing algorithm. The bins are constituted based on the temperature of the cores $\{T_1, \ldots, T_n\}$ using the density-based spatial clustering of applications with noise (DBSCAN) clustering method (*line 1 in Algorithm 1*) [40]. Density-Based Spatial Clustering of Applications with Noise (DBSCAN) is a clustering algorithm commonly used in data mining and machine learning. It is particularly useful for identifying clusters of data points in a dataset, especially when the data points are not easily separable using traditional clustering methods. The DBSCAN algorithm works by grouping data points that are close to each other in a high-density region. this algorithm defines a parameter called the

---

**Algorithm 2:** Task-to-Bin Assignment

---

**Input**: Formed bins $\{B_b, b \in \{1, \ldots, n_{Bin}\}\}$, set of the temperature of bins $\{T_{B_b}, b \in \{1, \ldots, n_{Bin}\}\}$, pair of task and its temperature $\{task, T_{task}\}$

**Output**: $\{task, selected \, bin\}$

**Parameters**: A set of temperatures of the bins $\{T_{B1}, \ldots, T_{Bn}\}$, $M_{dist(i,j)}$

1:   $Lb$ = List of bins for a task
2:   **for** i in range Bins number
3:     **if** $T_{Bi}$ is close to $T_{task}$
4:      add bin to $Lb$
5:   **if** $|Lb| \geq 2$
6:     **for** every bin in the $Lb$
7:      calculate $M_{dist(i,j)}$
8:      choose min ($M_{dist(i,j)}$)
9:   **return** $\{task, selected \, bin\}$

---

"*Epsilon*" value, which specifies the maximum distance between two points for them to be considered part of the same cluster. Another parameter called "*m*" specifies the minimum number of data points required to form a cluster. This algorithm starts by randomly selecting a data point and identifying all other points within the epsilon distance. If there are at least "*m*" data points within the epsilon distance, then a new cluster is formed. The algorithm then continues to add nearby data points to the cluster until no more points can be added. Any remaining data points that do not belong to any cluster are considered noise. One of the benefits of DBSCAN is its ability to handle clusters of varying shapes and densities. It is also robust to outliers and noise, which can be difficult to handle using other clustering methods. In our proposed method, we leveraged the DBSCAN method to form bins based on the temperature of cores, where *Epsilon* and *m* are the maximum difference between cores temperatures in each bin and the minimum number of cores that should be in each bin, respectively [41] (*line 5 in Algorithm 1*). Note that the value of *Epsilon* is obtained empirically. In this work, based on the simulation results (see Section V, subsection *B*), we found that $Epsilon = 0.7$ provides a good set of results. Also, *M* is the minimum number of points that are required to form a bin, i.e., the minimum number of cores that should be in each bin. As mentioned before, the DBSCAN method may consider the cores whose temperature differences with others are more than Epsilon, as noise (*line 7 in Algorithm 1*). In this work, we consider *M=1*. Thus, the DBSCAN method may also form some bins composed of a single core whose temperature is very different than the other cores. Finally, the output of this step is the formed bins $\{B_b, b \in \{1, \ldots, n_{Bin}\}\}$ that include the list of the cores in each bin (*lines 9-11 in Algorithm 1*).

**Task-to-Bin Assignment:**
After the bin packing step, bins are sorted by their average temperatures. To reduce the thermal variation (see Section

III subsection *B*) and to achieve a higher MTTF$_{TC}$, each task is assigned to a bin whose temperature (i.e., the average temperature of cores inside the bin) is the closest one to that of the task. Algorithm 2 represents the proposed task-to-bin assignment step. The formed bins $\{B_b, b \in \{1, \ldots, n_{Bin}\}\}$, the set of the temperature of bins $\{T_{B_b}, b \in \{1, \ldots, n_{Bin}\}\}$, and the pair of the arrived task and its temperature $\{task, T_{task}\}$ are the inputs to this algorithm. The output is the pair of the task and its selected bin $\{task, selected \, bin\}$. $M_{dist(i,j)}$ is the distance between two cores (core *i* and core *j*) in the manycore system. For the arrived task, the bin with the closest average temperature will be chosen (*line 3 in Algorithm 2*).

Note, if there is more than one bin that can be selected to assign the task, to achieve higher performance, the bin with a lower Manhattan distance ($M_{dist}$) to the previous bin that the related task was mapped, is chosen (*lines 5-8 in Algorithm 2*). As noted in [1], optimizing communication overheads in multi-threaded applications often involves concentrating the application's threads in the closest possible region. By minimizing the Manhattan distance between threads, we can effectively reduce the overall communication overhead and improve system performance. Considering this aspect, in [1] the performance improved up to 17.5% compared to the baseline. When there are multiple bins available to map a task, i.e., the average temperature of those bins is appropriate for mapping the arrived task, we use the Manhattan distance between the center core of each bin and the core we assigned the previous task. The Manhattan distance is a distance metric that considers the absolute differences between the coordinates of two points in space and is commonly used in computer science applications. We choose the bin with the shortest Manhattan distance to the core we assigned the previous task to. This is done to avoid a performance overhead.

---

**Algorithm 3:** Core Selection Inside the Bin

---

**Input**: A *task*, Set of free cores in the selected bin $\{C_{f1}, \ldots, C_{fn}\}$, Set of frequencies of cores $\{F_1, \ldots, F_n\}$, frequency requirements of task $F_{t1}$

**Output**: Pair of assigned task to the core $\{task, core\}$

**Parameters**: $M_{dist(i,j)}$

1: Sort cores of each bin, based on their frequency
2: **if** task is from high performance application
3:  select the core with highest frequency
4: **else**
5:  select the core with lowest frequency
6:   **if** |selected cores| $\geq 2$
7:   **for** every selected core
8:    calculate $M_{dist(i,j)}$
9:   choose min($M_{dist(i,j)}$)
10:  **return** $\{task, core\}$

---

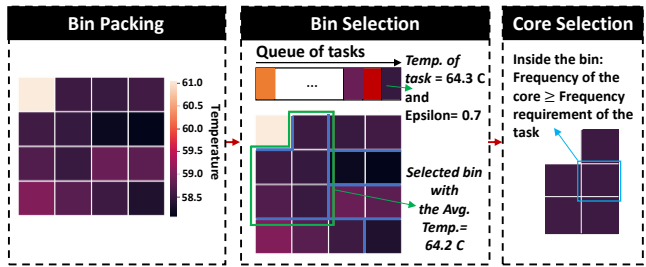

**FIGURE 7.** A real example for mapping the tasks of 64k points FFT application from SPLASH 2 benchmarks on a 16-cores system, including bin packing, bin selection, and core selection steps.

### 4.3.3 Core selection:

After choosing the bin, the task is mapped to a core considering the process variation information, i.e., the task of high-performance (low power) applications is mapped to the core with a higher (lower) operating frequency (power consumption). Algorithm 3 shows the proposed core selection step. A given task, a set of free cores in the selected bin ($\{C_{f1}, ..., C_{fn}\}$) and their frequencies $\{F_1, ..., F_n\}$, and the task frequency constraint ($F_{t1}$) are the inputs of this step. In our approach, we adhere to the methodology described in the LifeGuard paper ([8]), which involves carrying out tasks on designated cores to measure their average power and temperature. Moreover, when determining the task performance requirements and frequency specifications, we have incorporated a methodology inspired by [9], which involves calculating the frequency requirement of threads in an application based on factors such as the execution time of threads, their deadlines, and the clock period. To execute the task-to-core assignment in our proposed technique, we rely on the frequency requirements of tasks. For this purpose, we utilize the following equation:

$$f_{req,p,q} = \frac{t_{execution}}{T_{clk} \times t_{d,p,q}} \qquad (10)$$

where $t_{execution}$ represents the execution time, $T_{clk}$ denotes the clock period, $t_{p,q}$ is $q^{th}$ task of $p^{th}$ application, and $t_{d,p,q}$ refers to the deadline of task $t_{p,q}$.

As a result, the pair $\{task, core\}$ is returned. Inside the selected bin, we search for the core whose frequency satisfies the task's performance or power requirements. For high-performance applications, the task is assigned to a free core with the highest frequency, while for low power applications, the task is assigned to a free core with the lowest power. *(lines 2-5 in Algorithm 3)*. In case there is more than one free core to assign the task, the core with a lower $M_{dist}$ to the previous core the related task was mapped, will be chosen *(lines 6-9 Algorithm 3)*. Also, if all the cores were busy, the task should wait in the tasks queue until a core finishes its task. FIGURE 7 shows a real example of these three steps, for mapping the tasks of 64k points FFT application from SPLASH 2 benchmarks on a 16-cores system. Note that in the bin selection step of this example, a task with a temperature of 64.3° C is assigned to the bin with an average temperature of 64.2° C.

## V. RESULTS AND DISCUSSION

First, we introduce the experimental setup followed by detailed evaluations of our proposed two-level task mapping and comparisons with the two state-of-the-art approaches, i.e., the thermal cycling-aware technique, the technique [1] presents, and the random task mapping technique.

### A. SIMULATION TOOLCHAIN

FIGURE 8 shows our simulation setup and tool flow for evaluating the proposed two-level mapping technique. As shown in this figure, we use the Snipersim simulator to execute the applications [42], in which the McPAT simulator was integrated [43]. We evaluated the manycore systems in 22 nm technology, with tiles the size of 0.7 mm × 0.8 mm. Also, the system configuration that we used is the gainestown configuration of SniperSim, which is based on Nehalem core configuration. Each tile includes a Nehalem core with a private L1 cache (size 256 kB) and a private L2 cache (size 512 kB). Note, the power output file generated by the McPAT should be converted to one of the inputs of the HotSpot simulator. By utilizing power traces, hotspot simulation can estimate the power dissipation and subsequent temperature rise of each core. These power traces capture the dynamic power behavior of individual cores over time, providing valuable insights into their power consumption patterns. Through thermal modeling and power-to-temperature conversion, the power dissipation data is transformed into temperature estimates, enabling a comprehensive analysis of the system's thermal behavior.

Afterward, the generated input file, the floorplan file of the simulated manycore architecture, and a HotSpot configuration file that includes the set of default thermal model parameters are fed to the HotSpot. To create the floorplan, the HotSpot comes with a floorplanning tool called HotFloorPlan. It takes a list of the functional blocks (components of the manycore system) and their characteristics, including areas of blocks, allowable aspect ratios, and the connectivity between the blocks, and then, generates the floorplan. Note that we used a fixed floorplan in 22-nm technology as the basis of our evaluations generated by the Hotfloorplan tool. Finally, the HopSpot has steady and transient temperatures as outputs [44]. These outputs are used to calculate the cores aging and MTTF. We conducted our experiments with a 16, 32, 64, and 256 cores systems The simulation results were obtained by running a random execution of a single mix of benchmark applications, which comprises 15 different applications of the SPLASH2 [14] and PARSEC [45] benchmark suites. Details of the evaluated application benchmarks are shown in TABLE 3The workload characteristics can have a significant impact on the thermal behavior and lifetime of a many-core system. TABLE 4 includes the $MTTF_{TC}$ results of running multiple mixes of benchmark applications on a 32 cores system. As the results show, and as expected, running a mix of 4 applications resulted in a higher MTTF compared to executing a mix of 10 applications since the lower workload of the system reduces the thermal density and temperature levels, which can improve the reliability and lifetime of the system. However, by running a mix of 4 applications our proposed method achieved less improvement rather than running mix of 6,8 or 10 applications. It demonstrates that our two-level thermal-cycling-aware technique can achieve more improvements on larger workloads. Based on the results, for the four investigated mix of applications (i.e., M1~M4), our proposed method achieved, on average, 31%, 20.6%, and 20.2% higher MTTF compared to the Random task-mapping, conventional TC- aware method, and the method proposed in [1], respectively. We also measured the execution time of the four considered mixed of applications including M1 (mixes of 4 apps), M2 (mixes of 6 apps), M3 (mixes of 8 apps), and M4 (mixes of 10 apps). Based on the results, applying our proposed method on the four mixes of benchmark applications resulted in, on average, 5.1% higher execution time compared to the normal execution of those mixes of applications.

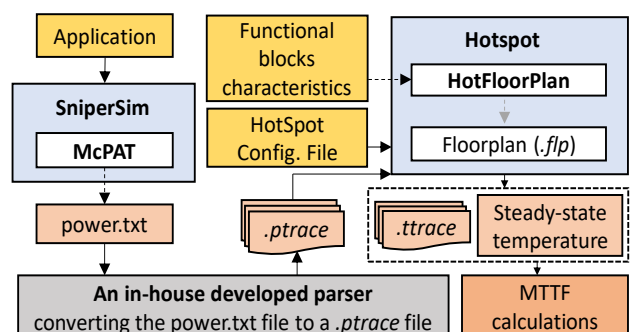

**FIGURE 8.** The simulation setup and tool flow for evaluations and computing the MTTF$_{TC}$ of the manycore system.

The arrival of execution requests is considered random such that we can capture the aging behavior of the system cores [8]. Similar to [1], thermal cycling model parameters introduced in (5) and (6) have been considered as follows: $E_{aTC} = 0.42eV$, $b = 2.35$, $T_{th} = 1\ C$, $K = 8.62 \times 10^{-5}\ eV/K$. Also, to calculate the $A_{TC}$, we considered $MTTF_{TC} = 10$ years with $\delta T = 20\ C$, $T_{max} = 70\ C$, $m = 10$, and 1-hour time duration.

Note that the proposed technique is a software-based solution and does not require any dedicated hardware. In our experiments, we implemented the proposed technique as a part of the operating system. In detail, our proposed algorithm is implemented as a Python script that runs as a separate process alongside the target applications. The space overhead of our Python script is relatively small, as it does not require any additional hardware. However, it requires some additional memory resources to save the cores' status, tasks' status, and bin information used by the thermal-aware bin packing algorithm. To minimize the memory overhead of our technique, we saved this information as simple look-up tables in text files, which can be a more memory-efficient way of storing the necessary information. However, the exact amount of required memory depends on the size of the task set, e.g., in our simulation, a memory with the size of 8.28 kB was used to store the required information. Note that the time complexity of the proposed technique is $O(nlogn)$, where n is the number of cores.

### B. EPSILON EFFECT ON THE MANYCORE SYSTEM'S MTTF

As mentioned before, Epsilon is the maximum allowed difference between temperatures of cores inside the composed bins. However, the appropriate Epsilon, in terms of its effect on the system $MTTF_{TC}$, is found empirically. Therefore, we examined the different values of *Epsilon* for the 16, 32, and 64-cores systems to perform the 64k points FFT application from SPLASH2 benchmark. The results of this investigation are shown in FIGURE 9. As the figure represents, for the different sizes of manycores, and *Epsilon* ranging from 0.1 to 0.7, the highest $MTTF_{TC}$ of the system is achieved when the *Epsilon* is 0.7. Thus, in our evaluations, we consider *Epsilon* 0.7.

TABLE 3
DETAILS OF THE EVALUATED APPLICATION BENCHMARKS

| | Memory-intensive | Compute-intensive | Mixed (memory and compute-intensive) |
|---|---|---|---|
| **SPLASH2** | FMM, Cholesky, LU | Raytrace, Radix | FFT, Radiosity, Barnes, Ocean |
| **PARSEC** | Dedup, VIPS, Swaptions | Black-Scholes, X264 | Canneal |

TABLE 4
THE AVERAGE $MTTF_{TC}$ OF A 32-CORE SYSTEM WHEN RUNNING MULTIPLE MIXES OF BENCHMARKS.

| Applications | Avg. $MTTF_{TC}$ (Year) | | | |
|---|---|---|---|---|
| | Random task-mapping | Conv. TC-aware method | Our proposed method | The method proposed in [1] |
| M1[1] | 6.5 | 6.94 | 8.19 | 6.96 |
| M2[2] | 5.73 | 6.25 | 7.57 | 6.28 |
| M3[3] | 5.53 | 6.09 | 7.28 | 6.11 |
| M4[4] | 4.83 | 5.36 | 6.65 | 5.38 |

1. A mix of FFT, Radiosity, FFM, and Raytrace
2. A mix of FFT, Radiosity, FFM, Raytrace, LU, and Radix
3. A mix of FFT, Radiosity, FFM, Raytrace, LU, Radix, Oceans, and Cholesky
4. A mix of FFT, Radiosity, FFM, Raytrace, LU, Radix, Oceans, Cholesky, Barnes, and X264

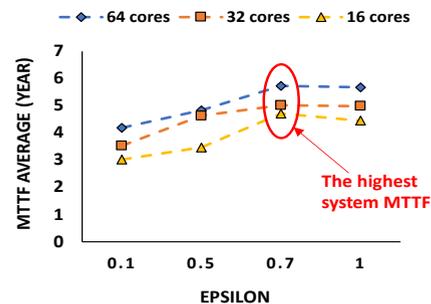

FIGURE 9. The $MTTF_{TC}$ of 16, 32, and 64-cores system, under executing the 64k points FFT application from SPLASH2 benchmark, for the different values of Epsilon

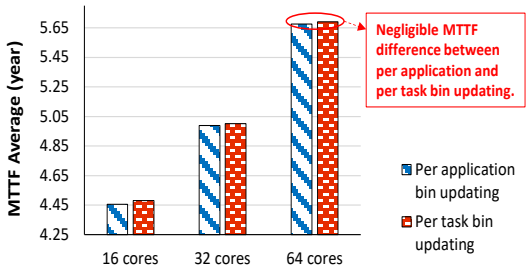

**FIGURE 10.** The comparison of updating the bins per task and per application for 16, 32, and 64-cores systems under executing nine ap-plications of SPLASH2.

### C. UPDATING THE BINS COMPOSITIONS

A factor that may affect the system overall performance, is updating the bins during multiple runs. Due to temperature changes in the cores, it is necessary to update the bin packing level after each run. However, this update process incurs time overhead, which may affect the overall performance of the system. The update can be performed per task or application, but the potential time overhead should be considered when designing the system architecture. illustrates the result of updating the bins, per task and application, for executing nine applications of SPLASH2, and for 16, 32, and 64 core systems. As the results show, per task update has negligible improvement. Therefore, to remove the timing overheads, it is preferred to update the bins per application. Compared to the execution time of the target application without bin packing, updating bins incurs an average execution time overhead of 3.3%.

### D. DIFFERENT AGING MECHANISMS IMPACT ON MTTF.

FIGURE 11 shows the average MTTF of the investigated task mapping algorithms, including random task mapping, conventional TC-aware task mapping, the proposed two-level task mapping, and the task mapping method proposed in [1], under the nine applications of SPLASH2 and six applications of PARSEC benchmark suites. Based 20%, 9.6%, and 25% MTTF improvement, respectively, compared to the other investigated task mapping algorithms.

### E. THERMAL CYCLING EFFECT

FIGURE 12 represents an example of the thermal cycling effect on the average $MTTF_{TC}$ of a 16-core system when executing the nine applications from the SPLASH2 benchmark. Note that in this exploration, no method has been applied to reduce the TC effects. Based on the results, more temperature variations (FIGURE 12. a) result in 54% lower $MTTF_{TC}$ compared to when there are no thermal variations (FIGURE 12. b), i.e., the thermal variation in lower temperature level may result in a lower $MTTF_{TC}$ than when the system operates in higher temperature level but without thermal variations (thermal cycling).

### F. RELIABILITY EVALUATION

In this subsection, we compare our proposed mapping algorithm with the two state-of-the-art techniques, the random and the conventional thermal-cycling-aware task mapping techniques in terms of the $MTTF_{TC}$ of the manycore systems, for the SPLASH2 and PARSEC benchmark suite applications. The studied task mapping techniques were investigated in two situations: when the number of tasks is equal to the number of cores, and when the number of tasks is more than the number of cores.

FIGURE 13 shows the simulation results for the first situation, where the average $MTTF_{TC}$ was obtained for 16, 32, and 64 cores systems. For the 16 cores system, the average $MTTF_{TC}$ improvement was 31% (9%) compared to the random (conventional TC-aware) task mapping technique. For the 32-core system, our proposed technique was achieved, on average, 53% and 20% higher $MTTF_{TC}$, compared to conventional random and conventional TC-aware task mapping techniques, respectively. These numbers for 64 cores system were 41% and 16%, respectively.

Based on the simulation results, for the 16 cores system, the average $MTTF_{TC}$ improvement of our proposed method was 19% compared to [1]. For the 32-core system, our proposed technique achieved, on average, 20% higher $MTTF_{TC}$, compared to [1]. This number for the 64 cores system was 15%.

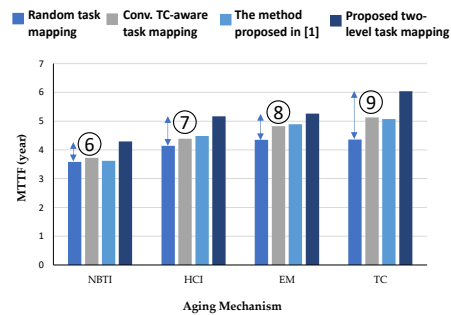

**FIGURE 11.** Average MTTF of the investigated task mapping algorithms for the different aging mechanisms. Key observations: ⑥, ⑦, ⑧, and ⑨: Our proposed method achieved up to 38% improvement in $MTTF_{TC}$ compared to the random task mapping method.

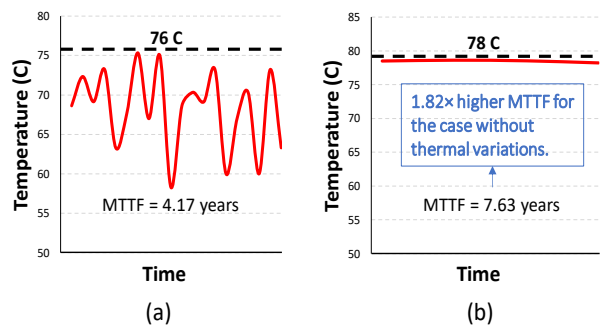

**FIGURE 12.** Thermal variations effect on the $MTTF_{TC}$ of a 16-cores system, when executing nine applications from SPLASH2 benchmark, when a) there are high thermal variations with the maximum temperature of 76° C, b) there are not thermal variations while the temperature.

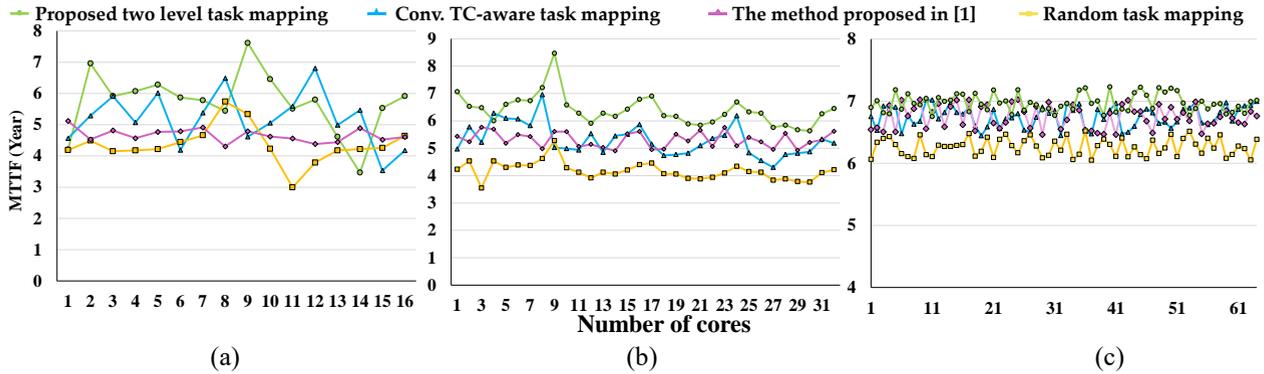

**FIGURE 13.** The average MTTF$_{TC}$ of the investigated manycore systems when the number of tasks is equal to the number of system cores, for a) 16, b) 32, and c) 64 core systems.

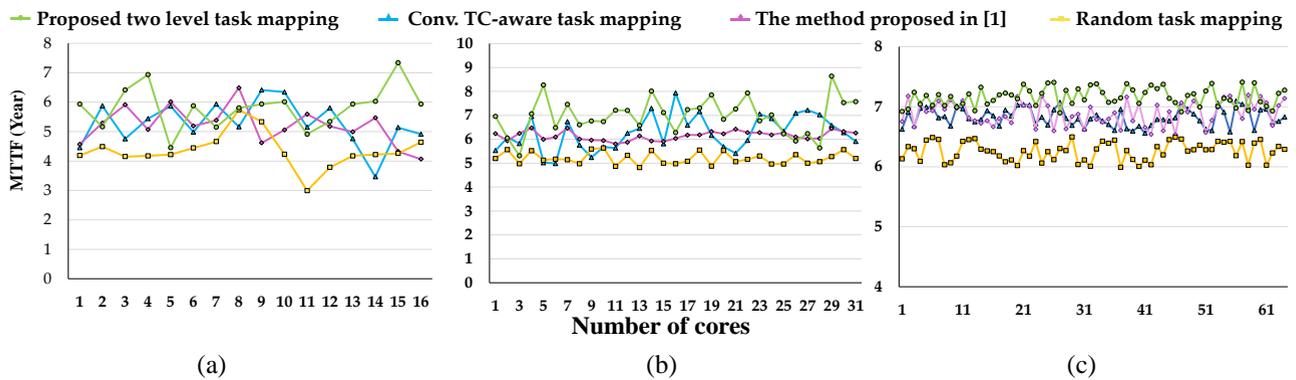

**FIGURE 14.** The average MTTF$_{TC}$ of the investigated systems when the number of tasks is more than the number of system cores, for a) 16, b) 32, and c) 64 core systems.

FIGURE 14 shows the average MTTF$_{TC}$ of the system for the second situation when the numbers of tasks are more than the cores. For the 16 (32, 64) cores system, our proposed two-level task mapping technique was achieved, on average, 23% (51%, 40%) and 4% (18%, 17%) higher MTTF$_{TC}$, compared to the random and conventional TC-aware task mapping techniques, respectively. Based on the results, for the 16, 32, and 64 cores systems, in both studied situations, our proposed mapping technique was achieved to, on average, 31 % and 18% higher MTTF$_{TC}$ compared to the conventional random and TC-based task mapping techniques, respectively. For the 16 (32, 64) cores system, our proposed two-level task mapping technique achieved, on average, 19.5% (21%, 13.8%) higher MTTF$_{TC}$, compared to [1].

In addition, we conducted experiments on smaller systems consisting of 2, 4, and 8 cores. The experimental results, presented in FIGURE 15, demonstrate an improvement of up to 4.3% in the MTTF$_{TC}$ of the system. FIGURE 15.a depicts the average Mean Time to Failure Time Cost (MTTF$_{TC}$) of the system in the first scenario, where the number of tasks is equal to the number of cores. For the 2 (4, 8) cores system, our proposed two-level task mapping technique achieved, on average, 0.2% (2%, 3%) and 0.1% (1%, 2.7%) higher MTTF$_{TC}$ than the random and conventional TC-aware task mapping techniques, respectively. Furthermore, compared to [1], our proposed technique achieved, on average, 0.1% (0.2%, 1%) higher MTTF$_{TC}$ for the 2 (4, 8) cores system. These findings are visualized on FIGURE 15.b, which shows the simulation results for the second scenario, where the average MTTF$_{TC}$ was obtained for 2, 4, and 8-core systems. In the 2-core system, our proposed technique demonstrated a 3% (1%) average MTTF$_{TC}$ improvement compared to the random (conventional TC-aware) task mapping technique. For the 4-core system, our proposed technique outperformed the conventional random and conventional TC-aware task mapping techniques with an average MTTF$_{TC}$ improvement of 4.1% and 2.8%, respectively. Similarly, in the 8-core system, the average MTTF$_{TC}$ improvement was 4.3% and 3.5%, respectively.

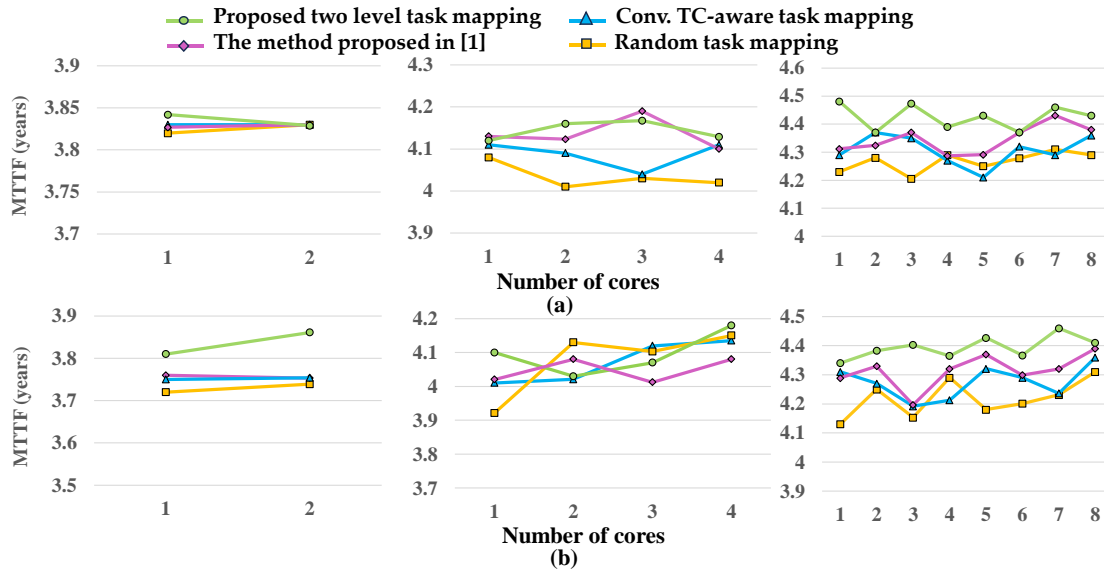

**FIGURE 15.** The average MTTF$_{TC}$ of the investigated systems when the number of tasks is more than the number of system cores, for 2, 4, and 8 core systems when a) numbers of cores are equal to number of tasks, and b) numbers of cores are less than number of tasks.

Based on the simulation results, our proposed method achieved a 1% average MTTF$_{TC}$ improvement compared to [1] in the 2-core system. For the 4-core system, our proposed technique achieved, on average, a 1.8% higher MTTF$_{TC}$ compared to [1]. In the 8-core system, this improvement reached 3.6%.

This indicates that our proposed method is more effective for systems with a higher number of cores. The underlying principle of our technique revolves around the efficient packing of cores into bins with similar temperatures. In smaller systems, with fewer available bins, the task mapping process may resemble traditional task mappings without the utilization of bins. However, as the number of cores increases, the advantages of our bin-based approach become more pronounced. The temperature-aware bin packing allows for better thermal management and improved reliability in larger systems, resulting in the observed reliability gains.

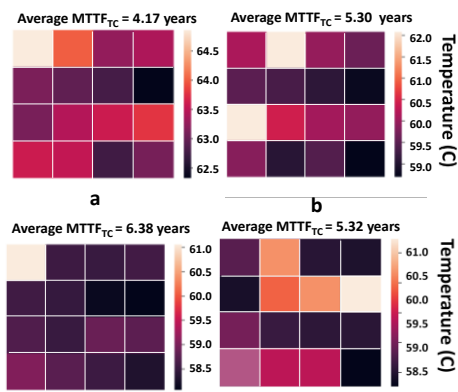

**FIGURE 16.** The heatmap of a 16-core system after executing nine applications of SPLASH2, and six applications of PARSEC benchmark suites, for a) random, b) conventional TC-aware, c) our pro-posed task mapping and d) [1] techniques when the Epsilon is 0.7.

FIGURE 16 shows the heatmap of the different studied task mapping techniques for a 16-core system after executing nine applications of SPLASH2 and six applications of PARSEC benchmark suites. Based on the results, our proposed two-level task mapping technique achieved 53% and 20% higher MTTF$_{TC}$ compared to the random and conventional TC-aware task mapping techniques, respectively. Based on the results, our proposed two-level task mapping technique achieved 20% higher MTTF$_{TC}$ compared to the [1].

## VI. CONCLUSION

This paper presents a two-level task mapping technique considering thermal cycling and process variation information, as well as the heat exchange effect among the adjacent cores. In the first level, based on the temperature of cores and tasks, bin packing and task-to-bin assignment steps are performed, where bins are composed based on the temperature of cores to reduce the thermal variations and heat exchange of adjacent cores. Afterward, in the second level, based on the task performance and power requirements, the most appropriate core in terms of frequency or power is selected to execute the task. The effectiveness of our proposed technique is evaluated on systems with 16, 32, 64, and 256 cores, using applications from the SPLASH2 and PARSEC benchmark suites. Simulation results show up to a 53% improvement in lifetime over the conventional random task mapping technique, with 20% and 21% improvements over the conventional TC-based technique and the method proposed in [1], respectively. The main limitation of our proposed algorithm is that it requires some information about tasks and cores that should be calculated at the design time, similar to some of the state-of-the-art TC-aware works such as [1]. As part of our ongoing research, we plan to explore methods to make our approach a runtime technique. Specifically, we aim to investigate machine learning algorithms for dynamically calculating the required information at run-time and adapting our approach accordingly.

## REFERENCES

[1] M. -H. Haghbayan, A. Miele, Z. Zou, H. Tenhunen and J. Plosila, "Thermal-Cycling-aware Dynamic Reliability Management in Many-Core System-on-Chip," in *proc. IEEE Design, Autom. & Test in Europe Conference & Exhibition (DATE)*, 2020, pp. 1229-1234.

[2] A. Iranfar, M. Kamal, A. Afzali-Kusha, M. Pedram and D. Atienza, "TheSPoT: Thermal Stress-Aware Power and Temperature Management for Multiprocessor Systems-on-Chip," *IEEE Transactions on Computer-Aided Design of Integrated Circuits and Systems*, vol. 37, no. 8, pp. 1532-1545, Aug. 2018.

[3] M. Kamal, A. Iranfar, A. Afzali-Kusha and M. Pedram, "A thermal stress-aware algorithm for power and temperature management of MPSoCs," in *proc. IEEE Design, Autom. & Test in Europe Conference & Exhibition (DATE)*, 2015, pp. 954-959.

[4] H. Haghbayan, A. Miele, O. Mutlu and J. Plosila, "Run-Time Resource Management in CMPs Handling Multiple Aging Mechanisms," in *IEEE Transactions on Computers*, vol. 72, no. 10, pp. 2872-2887, Oct. 2023.

[5] A. Iranfar, S. N. Shahsavani, M. Kamal and A. Afzali-Kusha, "A heuristic machine learning-based algorithm for power and thermal management of heterogeneous MPSoCs," in *Proc. IEEE/ACM ISLPED*, 2015, pp. 291-296.

[6] I. Ukhov, M. Bao, P. Eles, and Z. Peng, "Steady-state dynamic temperature analysis and reliability optimization for embedded multiprocessor systems," in *Proc. 49th Annual Design Automation Conference (DAC)*, Jun. 2012, pp. 197–204.

[7] G. Martinez, R. Sandoval-Arechiga, LO. Solis-Sanchez, and others, "A Survey of MPSoC Management toward Self-Awareness". *Micromachines.* 2024;

[8] V. Rathore, V. Chaturvedi, A. K. Singh, T. Srikanthan and M. Shafique, "Life Guard: A Reinforcement Learning-Based Task Mapping Strategy for Performance-Centric Aging Management," *56th Design Automation Conference (DAC)*, 2019, pp. 1-6.

[9] V. Rathore, V. Chaturvedi, A. K. Singh, T. Srikanthan and M. Shafique, "Longevity Framework: Leveraging Online Integrated Aging-Aware Hierarchical Mapping and VF-Selection for Lifetime Reliability Optimization in Manycore Processors," *IEEE Trans. on Computers*, vol. 70, no. 7, pp. 1106-1119, 1 July 2021.

[10] H. Khdr, M. Shafique, S. Pagani, A. Herkersdorf and J. Henkel, "Combinatorial Auctions for Temperature-Constrained Resource Management in Manycores," *IEEE Transactions on Parallel and Distributed Systems*, vol. 31, no. 7, pp. 1605-1620, 1 July 2020.

[11] S. M. P. Dinakarrao, et al., "Application and thermal-reliability-aware reinforcement learning-based multicore power management," *ACM J. Emerg. Tech. Comput. Syst.*, vol. 15, no. 4, pp. 33, Dec. 2019.

[12] D. Gnad, et al., "Hayat: Harnessing Dark Silicon and variability for aging deceleration and balancing," *52nd Design Automation Conference (DAC)*, 2015, pp. 1-6.

[13] A. Silva, I. Weber, A. L. d. M. Martins, and F. G. Moraes, "Dynamic Thermal Management in Many-Core Systems Leveraged by Abstract Modeling," *IEEE ISCAS*, 2021, pp. 1-5.

[14] S. C. Woo, et. al., "The SPLASH-2 programs: characterization and methodological considerations," *Proc. 22nd Annual International Symposium on Computer Architecture, 1995*, pp. 24-36.

[15] B. Pourmohseni, S. Wildermann, F. Smirnov, P. E. Meyer and J. Teich, "Task Migration Policy for Thermal-Aware Dynamic Performance Optimization in Many-Core Systems," in *IEEE Access*, vol. 10, pp. 33787-33802, 2022.

[16] Gaurav Narang, Aryan Deshwal, Raid Ayoub, Michael Kishinevsky, Janardhan Rao Doppa, and Partha Pratim Pande. 2023. "Dynamic Power Management in Large Manycore Systems: A Learning-to-Search Framework". *ACM Transactions on Design Automation of Electronic Systems* 28, 5, Article 84 (September 2023).

[17] M.S. Mohammed., A.A Al-Kubati, N. Paraman, A.A.H Ab Rahman, and M.N. Marsono, "DTaPO: Dynamic thermal-aware performance optimization for dark silicon many-core systems," *Electronics*, 9(11), p.1980.

[18] M.S. Mohammed, A. Al-Dhamari, M. Hamdan, A.M.H. Saad, A.S. Abdul-Qawy, and M.N. Marsono, '3D-DNaPE: Dynamic Neighbor-Aware Performance Enhancement for Thermally Constrained 3D Many-Core Systems," *IEEE Access*, 11, pp.131964-131978.

[19] Wenjun Lin, Weiwei Lin, Jianpeng Lin, Haocheng Zhong, Jiangtao Wang, Ligang He, "A multi-agent reinforcement learning-based method for server energy efficiency optimization combining DVFS and dynamic fan control", *Sustainable Computing: Informatics and Systems*, Volume 42, 2024, 100977, ISSN 2210-5379,

[20] V. Rathore et al., "HiMap: A hierarchical mapping approach for enhancing lifetime reliability of dark silicon manycore systems," in *proc. Design, Automation & Test in Europe Conference & Exhibition (DATE)*, 2018, pp. 991-996.

[21] S. Rahimipour, et al., "Low-Power, Highly Reliable Dynamic Thermal Management by Exploiting Approximate Computing," *IEEE Transactions on Very Large-Scale Integration (VLSI) Systems*, vol. 28, no. 10, pp. 2210-2222, Oct. 2020.

[22] A. Das, et al., "Reinforcement Learning-Based Inter- and Intra-Application Thermal Optimization for Lifetime Improvement of Multicore Systems," in *Proc. Design Autom. Conf. (DAC)*, 2014, pp. 170:1–170:6.

[23] J. Saber-Latibari, et al., "READY: Reliability- and Deadline-Aware Power-Budgeting for Heterogeneous Multicore Systems," *IEEE Transactions on Computer-Aided Design of Circuits and Systems*, vol. 40, no. 4, pp. 646-654, April 2021.

[24] M. Kamal, A. Iranfar, A. Afzali-Kusha, and M. Pedram, "A thermal stress-aware algorithm for power and temperature management of MPSoCs," in *proc. Design, Automation & Test in Europe Conference & Exhibition (DATE)*, 2015, pp. 954-959.

[25] T. Chantem, Y. Xiang, X. S. Hu and R. P. Dick, "Enhancing multicore reliability through wear compensation in online assignment and scheduling," in *Proc. Design, Automation & Test in Europe Conference & Exhibition (DATE)*, 2013, pp. 1373-1378.

[26] A. Baldassari, C. Bolchini and A. Miele, "A dynamic reliability management framework for heterogeneous multicore systems," *IEEE DFT*, 2017, pp. 1-6.

[27] J. Srinivasan, S. V. Adve, P. Bose, and J. A. Rivers, "The Case for Lifetime Reliability-Aware Microprocessors," in *Proc. of Intl. Symp. on Computer Architecture (ISCA)*, 2004, pp. 276–287.

[28] Y. Ma, J. Zhou, T. Chantem, R. P. Dick, S. Wang, and S. X. Hu, "Improving Reliability of Soft Real-Time Embedded Systems on Integrated CPU and GPU Platforms," in *IEEE TCAD*, vol. 39, no. 10, pp. 2218–2229, 2020.

[29] Shounak Chakraborty, Yanshul Sharma, and Sanjay Moulik. "TREAFET: Temperature-Aware Real-Time Task Scheduling for FinFET based Multicores," *ACM Transactions on Embedded Computer System.* 23, 4, Article 61 (July 2024).

[30] S. Moulik and Z. Das, "TASOR: A Temperature-Aware Semi-Partitioned Real-time Scheduler," *TENCON* 2019 - 2019 IEEE Region 10 Conference (TENCON), Kochi, India, 2019, pp. 1578-1583.

[31] Y. Sharma, S. Moulik and S. Chakraborty, "RESTORE: Real-Time Task Scheduling on a Temperature Aware FinFET based Multicore," *2022 Design, Automation & Test in Europe Conference & Exhibition (DATE)*, Antwerp, Belgium, 2022, pp. 608-611.

[32] Sharma, Yanshul & Moulik, Sanjay, "RT-SEAT: A hybrid approach based real-time scheduler for energy and temperature efficient

heterogeneous multicore platforms," *Results in Engineering*. 16. 100708. 10.1016/j.rineng.2022.100708.

[33] S. Moulik, Z. Das and G. Saikia, "CEAT: A Cluster based Energy Aware Scheduler for Real-Time Heterogeneous Systems," *IEEE International Conference on Systems*, Man, and Cybernetics (SMC), Toronto, ON, Canada, 2020, pp. 1815-1821,

[34] Y. Sharma and S. Moulik. "FATS-2TC: A Fault Tolerant Real-time Scheduler for energy and temperature aware heterogeneous platforms with Two types of Cores," *Microprocess*. Microsyst. 96, C (Feb 2023).

[35] Y. Sharma and S. Moulik. "CETAS: a cluster-based energy and temperature efficient real-time scheduler for heterogeneous platforms," In *Proc of the 37th ACM/SIGAPP Symposium on Applied Computing (SAC '22)*. Association for Computing Machinery, New York, NY, USA, 501–509.

[36] Akhil Langer, et al. "Energy-efficient computing for HPC workloads on heterogeneous manycore chips," in *proc. the Sixth International Workshop on Programming Models and Applications fo Multicores and Manycores. (PMAM '15)*, 2015.

[37] SD. Dowining, and D.F. FSocie, "Simple rainflow counting algorithm," *International Journal of Fatigue*, 4(1), pp. 31-40, 1982.

[38] J. Kong et al., "Recent thermal management techniques for microprocessors", *ACM Comput. Surv.*, vol. 44, 2012.

[39] J.W. McPherson, "Reliability Physics and Engineering Time-to-Failure Modeling", 2nd edition, *Springer*, USA, 2013.

[40] K. Khan, et al., "DBSCAN: Past, present, and future," *The Fifth International Conference on the Applications of Digital Information and Web Technologies (ICADIWT 2014)*, 2014, pp. 232-238.

[41] M. Ester, H. P. Krigel, J. Sander, and X. Xu, "A Density-Based Algorithm for Discovering Clusters in Large Spatial Databases with Noise," *Proc. 2nd International Conference on Knowledge Discovery and DataMining, Portland, WA*, 1996, pp. 226-231.

[42] T. E. Carlson, W. Heirman and L. Eeckhout, "Sniper: Exploring the level of abstraction for scalable and accurate parallel multicore simulation," *Proc. International Conference for High-Performance Computing, Networking, Storage and Analysis*, 2011, pp. 1-12.

[43] S. Li, J. H. Ahn, R. D. Strong, J. B. Brockman, D. M. Tullsen and N. P. Jouppi, "McPAT: An integrated power, area, and timing modeling framework for multicore and manycore architectures," *42nd Annual IEEE/ACM MICRO*, 2009, pp. 469-480.

[44] W. Huang, et. al., "HotSpot: A compact thermal modeling methodology for early-stage VLSI design," *IEEE Trans. Very Large Scale Integr. (VLSI) Syst.*, vol. 14, no. 5, pp. 501–513, May 2006.

[45] C. Bienia *et al.*, "The PARSEC benchmark suite: Characterization and architectural implications," in *PACT*, 2008.

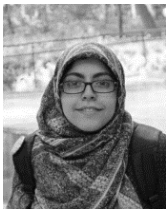
**Fatemeh Hossein-Khani** received her B.Sc. degree from Amirkabir University of technology, Tehran, Iran, and M.Sc. degree from the Tarbiat Modares University, Tehran, Iran, both in Computer Engineering, in 2017 and 2022, respectively. Her research interests include fault-tolerant system design, reconfigurable system, AI and machine learning hardware and system-level design, and hardware accelerators.

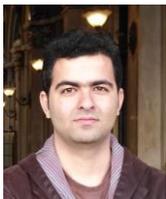
**Omid Akbari** received the B.Sc. degree from the University of Guilan, Rasht, Iran, in 2011, the M.Sc. degree from Iran University of Science and Technology, Tehran, Iran, in 2013, and the Ph.D. degree from the University of Tehran, Iran, in 2018, all in Electrical Engineering, Electronics - Digital Systems sub-discipline. He was a visiting researcher in the CARE-Tech Lab. at Vienna University of Technology (TU Wien), Austria, from Apr. to Oct. 2017, and a visiting research fellow under the Future Talent Guest Stay program at Technische Universität Darmstadt (TU Darmstadt), Germany, from Jul. to Sep. 2022. He is currently an assistant professor of Electrical and Computer Engineering at Tarbiat Modares University, Tehran, Iran, where he is also the Director of the Computer Architecture and Dependable Systems Laboratory (CADS-Lab). His current research interests include embedded machine learning, reconfigurable computing, energy-efficient computing, distributed learning, and fault-tolerant system design.

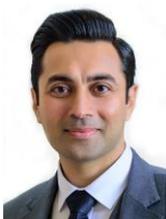
**Muhammad Shafique** (Senior Member, IEEE) received the Ph.D. degree in computer science from the Karlsruhe Institute of Technology (KIT), Germany, in 2011. In Oct.2016, he joined the Institute of Computer Engineering at the Faculty of Informatics, Technische Universität Wien (TU Wien), Vienna, Austria as a Full Professor of Computer Architecture and Robust, Energy-Efficient Technologies. Since Sep.2020, Dr. Shafique is with the New York University (NYU), where he is currently a Full Professor and the director of eBrain Lab at the NYU-Abu Dhabi in UAE, and a Global Network Professor at the Tandon School of Engineering, NYU-New York City in USA. He is also a Co-PI/Investigator in multiple NYUAD Centers. His research interests are in AI & machine learning hardware and system-level design, brain-inspired computing, quantum machine learning, cognitive autonomous systems, wearable healthcare, energy-efficient systems, robust computing, hardware security, emerging technologies, FPGAs, MPSoCs, and embedded systems.